\documentstyle[12pt]{article}
\def\be{\begin{equation}}
\def\bd{\begin{displaymath}}
\def\ee{\end{equation}}
\def\ed{\end{displaymath}}
\begin{document}

\hfill SNUTP-93-40

\hfill July 1993

\vskip 1.5cm
\begin{center}
{\bf {\huge Hadronic axion, baryogenesis and dark matter}}
\vskip 2cm
\Large{Bengt-{\AA}ke Lindholm
\footnote{Supported by {\it The Center for Theoretical Physics,
Seoul National University} and by {\it The Swedish Institute}} \\
\vskip .8cm
Center for Theoretical Physics\\
Seoul National University\\
Seoul 151-742}\\
Korea
\end{center}
\vskip 2.0cm
{\center Abstract\\}
\vskip .5cm
We investigate the prospects of an extra global symmetry, added to the
Standard Model in the context of baryogenesis. The $PQ$ symmetry is studied
as an example. We show that the hadronic axion provides a protection of the
B-asymmetry even if $B-L$=0 and that there are no limits on the neutrino
masses. It is also shown that there is a crucial difference between
the two possible invisible axion models.
A connection between the baryon asymmetry and dark matter is also pointed
out.
\newpage

\section{Introduction}
It has been known for a long time that $B+L$ symmetry is anomalous and
that  $B-L$ symmetry is anomaly free in the Standard Model. However, the
operator induced by the anomaly has negligible effect at zero temperature
\cite{t'Hooft}.
Some years ago the sphaleron solution to the classical $SU(2)$ gauge-Higgs
system was found \cite{mk}. It is a saddle point configuration and is
unstable. Soon after this \cite{krs}, it was realized that the sphaleron
mediates $B+L$  violating
interactions at a substantial rate at high temperatures.
Since then, this circumstance has been under intense investigation.
Recent reviews on this topic are given in refs. \cite{bau-review}.
Contrary to earlier beliefs it was shown in refs. \cite{krs} and \cite{ht}
that a $B$ asymmetry created
at GUT scale not necessarily is washed out by the sphalerons. Their
conclusions were that the $B$ asymmetry is proportional to a $B-L$
asymmetry that could have been created at GUT scale.
However, this immediately raises the question of  $B-L$ violating operators.
Operators of dimension five or higher coming from sources beyond the Standard
Model are posing a threat to the  $B-L$ asymmetry are they in equilibrium.
These issues have been investigated in refs. \cite{ht}, \cite{fglp} and
\cite{L-op}. The lowest
dimension operator that violates $B$ or $L$ is
the following $\Delta$L=2 dimension-five operator:
\be  \label{L-violating op}
\frac{m_{\nu}}{\it v^{2}}({\it l_{L}}H)^{2}
\ee
$l_{L}$ and $H$ are the left-handed lepton doublet and Higgs doublet
respectively.
Below the electroweak symmetry breaking scale this operator induces a
majorana mass for the neutrino. It also mediates different scattering
processes all of which break $L$-conservation.
The requirement that this operator be out of equilibrium, so that  $B-L$
asymmetry
is not washed out, puts limits on the neutrino mass and in ref. \cite{ht} the
following limit was derived:
\be  \label{4eV-limit}
m_{\nu} < \frac{4 eV}{(T_{B-L}/10^{10} GeV)^{1/2}}
\ee
where $T_{B-L}$ is the temperature at which the  $B-L$ asymmetry
is generated. This was elaborated further in ref. \cite{fglp} where decays of
heavy right-handed neutrinos also were considered. In that case the limit
is much stronger:
\be  \label{millieV-limit}
m_{\nu} < 10^{-3} eV
\ee
It has been pointed out in ref. \cite{iq} that there are ways to
circumvent this limit. Above a certain temperature in the minimal
supersymmetric Standard Model there are two more global symmetries that will
protect a primordial  $B-L$ asymmetry. These symmetries have $SU(2)$ and
$SU(3)$ anomalies. The constraint on the neutrino mass
coming from (\ref{L-violating op}) is in this case relaxed by four
orders of magnitude and reads:
\be  \label{10eV-limit}
m_{\nu} < 10 eV
\ee
Investigations along the line of extra global symmetries have also been
pursued in ref. \cite{extra global symm}. In that paper one more unbroken
global symmetry with an $SU(2)$-anomaly was introduced. Several new fields as
compared to the Standard Model were also introduced. It will protect a
primordial $B$
asymmetry even if  $B-L$=0. At the same time there is also the possibility
to have an asymmetry in the new charge that could be responsible for the
Dark Matter.
\section{An extra global symmetry}
In this paper we examine the prospects of the $PQ$ symmetry \cite{pq}
in protecting a primordial $B$ asymmetry. This extra symmetry has no $SU(2)$
anomaly which is a difference compared to the previously studied extra global
symmetries that can protect a primordial $B$ asymmetry.
Our main interest is the hadronic axion model \cite{ksvz}. The DFSZ-axion
\cite{dfsz} cannot protect a primordial $B$ asymmetry as will be shown. To be
more explicit, the model to be studied is the Standard Model with $N$
generations of quarks and leptons and one Higgs doublet. In order to
implement the hadronic
axion we also introduce one extra $SU(2)$-singlet quark with electric charge
$q$ and a complex $SU(2)$-singlet scalar.

\noindent We will calculate the $B$ and $L$ number in equilibrium in the
early Universe.
We assign a chemical potential to each of the
$N$ generations of quarks and leptons, to the $SU(3)\times SU(2)\times U(1)$
gauge bosons, to the complex Higgs doublet, to one complex scalar singlet
and to one $SU(2)$-singlet heavy quark. Here we will use the notation of ref.
\cite{ht} and the chemical potentials are assigned as follows: $\mu_{W}$ for
$W^{-}$, $\mu_{0}$ for $\phi^{0}$, $\mu_{-}$ for $\phi^{-}$, $\mu_{uL}$
for all
the left handed up-quark fields, $\mu_{uR}$ for all the right handed up-quark
fields,  $\mu_{dL}$ for all the left handed down-quark fields, $\mu_{dR}$ for
all the right handed down-quark fields. $\mu_{iL}$ for all the left handed
charged leptons fields, $\mu_{iR}$ for all the right handed charged lepton
fields. $\mu_{i}$ for all the left handed neutrino fields, $i=1$ to $N$.
$\mu_{\sigma}$
for the complex singlet scalar field and $\mu_{QL}$, $\mu_{QR}$ for the
left and right handed singlet heavy quarks.
Rapid interactions in the early Universe enforce the following equilibrium
relations among the chemical potentials:
\begin{eqnarray}   \label{interactions}
\mu_{W}&=&\mu_{-}+\mu_{0}, \hspace{3mm}
W^{-} \leftrightarrow \phi^{0}+\phi^{-} \nonumber  \\
\mu_{dL}&=&\mu_{uL}+\mu_{W}, \hspace{3mm}
W^{-} \leftrightarrow {\bar u}_{L}+d_{L} \nonumber  \\
\mu_{iL}&=&\mu_{i}+\mu_{W}, \hspace{3mm}
W^{-} \leftrightarrow {\bar \nu}_{iL}+e_{iL} \nonumber  \\
\mu_{uR}&=&\mu_{0}+\mu_{uL}, \hspace{3mm}
\phi^{0}  \leftrightarrow {\bar u}_{L}+u_{R} \nonumber \\
\mu_{dR}&=&-\mu_{0}+\mu_{dL}, \hspace{3mm}
\phi^{0}  \leftrightarrow {\bar d}_{R}+d_{L} \nonumber  \\
\mu_{iR}&=&-\mu_{0}+\mu_{iL}, \hspace{3mm}
\phi^{0}  \leftrightarrow {\bar e}_{iR}+e_{iL} \nonumber \\
\mu_{QR}&=&\mu_{QL}+\mu_{\sigma}, \hspace{3mm}
\sigma  \leftrightarrow \bar{Q}_{L}+Q_{R}
\end{eqnarray}
At temperatures high above the weak scale we can assume that all the chemical
potentials for the particles in the same $SU(3) \times SU(2) \times U(1)$
multiplet
are the same and hence all the corresponding gauge bosons have vanishing
chemical potential.
Moreover, due to cabbibo mixing the chemical potentials for the different
quark flavours are the same, whereas the chemical potentials for the
leptons are in general different.

\noindent
There are $10 + 3N$ chemical potentials and $5 + 2N$ relations. This gives us
$5 + N$ independent chemical potentials chosen to be $\mu_{uL}$, $\mu_{0}$,
$\mu_{W}$, $\mu_{QL}$, $\mu_{QR}$ and $\mu_{i}$ which corresponds to the
conserved charges:
$Q$, $T_{3}$, $PQ$, $B$, $B_{Q}$ and $L_{i}$. $B_{Q}$ is the fermion number
carried by the extra $SU(2)$-singlet quark.
The $SU(2)$ anomaly induces the following operator:
\be \label{anomaly op}
(q_{L}q_{L}q_{L}l_{L})^{N_{g}}
\ee
where $q_{L}$ and $l_{L}$ are the left-handed doublets.  $N_{g}$ is the
number of generations.
This will give us the following relation between the chemical potentials:
\be  \label{B+L-op}
N(\mu_{uL}+2\mu_{dL}) + \sum \mu_{i} =0
\ee
The number of chemical potentials is reduced by one and $B$ and $L_{i}$
are no longer separately conserved but instead we have $B-\sum L_{i}$
conserved\footnote{Strictly speaking there are $N$ conserved charges:
$\frac{1}{N}B-L_{i}$}.
The $PQ$-symmetry has an SU(3) anomaly which induces the following operator:
\be  \label{PQ-anomaly}
(q_{L}q_{L}(u_{R})^{c}(d_{R})^{c})^{N_{g}}(Q_{L}(Q_{R})^{c})
\ee
$c$ is charge conjugation.
Whether this operator gives rise to fast enough $PQ$-breaking processes at
finite temperature is still an open question.
It is of course tempting to assume the existence of QCD-sphalerons which
would mediate transitions in a similar way as the $SU(2)$-sphalerons mediate
$B+L$ violation in the weak sector.
It seems established that the rate of $B+L$
violating processes in the symmetric phase of the electroweak sector are
non-zero \cite{symmetricphase}.
In ref. \cite{mclerran} it is assumed that the $SU(2)$ sector, in the
symmetric
phase, in all essentials are the same as QCD. For this reason the rate for
the processes induced by (\ref{PQ-anomaly}) should also be non-zero.
However, there are reasons to be a little bit careful here since the
sphaleron is a solution of $SU(2)$-gauge-Higgs system in the broken phase.
Although there are indications
that the Higgs field decouples in the symmetric phase \cite{Higgs
decoupling}
of the $SU(2)$ sector there are still uncertainties about the roll of
the Higgses in the symmetric phase \cite{uncertainties}. We will in this
paper assume that the operator in (\ref{PQ-anomaly}) is out of equilibrium.

\noindent
In the relativistic limit the number densities are related to the
chemical potentials as \cite{kt}\footnote{We will in this paper neglect
mass effects.}:
\begin{eqnarray}  \label{n-density}
\frac{n_{+}-n_{-}}{s} &=& \frac{15g}{4{\pi}^{2} g_{*}}\frac{\mu}{T}
\hspace{1cm} for \hspace{2mm}fermions  \nonumber \\
\frac{n_{+}-n_{-}}{s} &=& \frac{15g}{2{\pi}^{2} g_{*}}\frac{\mu}{T}
\hspace{1cm} for \hspace{2mm}bosons
\end{eqnarray}
$n_{+}$ and $n_{-}$ are the number densities for particles and
anti-particles respectively. $g$ and $g_{*}$ are respectively the number of
internal and total degrees of freedom. $s$ is the entropy density.
We now have eqs. (\ref{interactions}), (\ref{B+L-op}),(\ref{n-density})
together with $\mu_{W}=0$ and the constraint of electric charge neutrality
of the Universe. From this we can
easily write down the different number densities as:
\begin{eqnarray}  \label{B,L}
B &=& \frac{15g}{4{\pi}^{2} g_{*}T}4N\mu_{uL}  \nonumber \\
L &=& \frac{15g}{4{\pi}^{2} g_{*}T}(-\frac{14N^{2}+9N}{2N+1}\mu_{uL}+
\frac{Nq}{2(2N+1)}(\mu_{QL}+\mu_{QR})) \nonumber \\
B_{Q} &=& \frac{15g}{4{\pi}^{2} g_{*}T}(\mu_{QL} +\mu_{QR}) \nonumber \\
PQ &=& \frac{15g}{4{\pi}^{2} g_{*}T}(2\mu_{\sigma}+
\frac{1}{2}(\mu_{QL}-\mu_{QR}))
\end{eqnarray}
As can be seen, the simple proportionality between $B$ and $B-L$  is
lost here. There are three conserved charges: $B-L$, $PQ$, and $B_{Q}$. We
can afford to break or initially put to zero two of them and still have one
chemical potential that is non-zero and proportional to the $B$ number.
This can be done in various ways.
$B_{Q}$ is conserved in the low energy sector, $PQ$ is spontaneously broken
at a scale $F_{a}$ and $B-L$ can be violated by higher dimension operators
from beyond the standard model.
\subsection{$B_{Q} \neq 0$}
Below the $PQ$ breaking scale $\mu_{\sigma}$ is equal to zero. From
eq.(\ref{interactions}) we then get $\mu_{QL} = \mu_{QR}$ which gives
\begin{eqnarray}  \label{B,L-2}
B &=& \frac{15g}{4{\pi}^{2} g_{*}T}4N\mu_{uL}  \nonumber \\
L &=& \frac{15g}{4{\pi}^{2} g_{*}T}(-\frac{14N^{2}+9N}{2N+1}\mu_{uL}+
\frac{Nq}{2N+1}\mu_{QL}) \nonumber \\
B_{Q} &=& \frac{15g}{4{\pi}^{2} g_{*}T}2\mu_{QL}
\end{eqnarray}
If the operator in (\ref{L-violating op}) is in
equilibrium there will be the extra condition $\mu_{0} = -\mu_{i}$
which will change $L$ and $B_{Q}$ to:
\begin{eqnarray}  \label{B,L-3}
L &=& -\frac{15g}{4{\pi}^{2} g_{*}T}12N\mu_{uL} \nonumber \\
B_{Q} &=& -\frac{15g}{4{\pi}^{2} g_{*}T}\frac{2}{q}(10N+3)\mu_{uL}
\end{eqnarray}
As can be seen,
in this case there is still one chemical potential, corresponding
to the $B_{Q}$ asymmetry. All the other charges can now be expressed in
terms of $B_{Q}$.
\begin{eqnarray}  \label{B,L-4}
B &=& -\frac{2Nq}{10N+3}B_{Q} \nonumber \\
L &=& \frac{6Nq}{10N+3}B_{Q} \nonumber \\
B-L &=& -\frac{8Nq}{10N+3}B_{Q} \nonumber \\
B+L &=& \frac{4Nq}{10N+3}B_{Q}
\end{eqnarray}
We do not need to worry about $L$ violating operators and of course there is
no limit on neutrino masses. $B_{Q}$ protects both $B$ and $B-L$ from
becoming zero. The exotic quark may make up part of the dark matter and in
this model there is a connection between $B_{Q}$ and $B$ which depend on the
charge of the exotic quark.
\subsection{$B_{Q}=0$}
{}From eq.(\ref{B,L}) we can see that $B_{Q}=0$ gives
$\mu_{QL} = -\mu_{QR}$. For $B$ and $L$, now only the terms proportional to
$\mu_{uL}$ in
eq.(\ref{B,L}) will survive. In this case we arrive at the same
expressions as derived in \cite{ht}:
\begin{eqnarray}  \label{B,L-5}
B &=& \frac{8N+4}{22N+13}(B-L) \nonumber \\
L &=& -\frac{14N+9}{22N+13}(B-L) \nonumber \\
B+L &=& -\frac{6N+5}{22N+13}(B-L)
\end{eqnarray}
This will be the case regardless of what happens with the $PQ$ symmetry.
With $B_{Q}=0$ it reduces to: $PQ \propto -\mu_{QL}$. This chemical
potential will not
show up in the expressions for $B$, $L$, $Q$ or the sphaleron condition.
The derivation of eq.(\ref{B,L-5}) will therefore go through unaffected of
whether $PQ$ is zero or not or whether the operator in (\ref{PQ-anomaly})
is in equilibrium or not. In this case the limit on neutrino masses from
eq.(\ref{millieV-limit}) will be valid.
{}From this we can see that the $DFSZ$-axion is unable to protect a primordial
$B$ asymmetry since the extra fields in that model will not show up in the
expressions for $B$, $L$, $Q$ or the sphaleron condition.

\section{Conclusions}
We have looked into effects of having extra global symmetries in the
Standard Model in the context of baryogenesis. As an example we have taken
the $PQ$ symmetry. If $B_{Q} \neq 0$ than the hadronic axion will provide a
protection for
a primordial $B$ asymmetry in much the same way as $B-L$ asymmetry can
protect a primordial $B$ asymmetry. At the same time all limits on the
neutrino masses are abolished. In this model there is also a connection
between dark matter abundance and $B$ asymmetry.
A difference between the two models of $PQ$  symmetry has been
shown. It turns out that the hadronic axion but not the DFSZ axion
protects the $B$ asymmetry.

\begin{center}
\large \bf Acknowledgement
\end{center}
I would like to thank Jihn E. Kim and S. Y. Choi for discussions during
the time this work was completed.
\bibliographystyle{unsrt}

\end{document}